\documentclass[prl,twocolumn,aps,amssymb,footinbib,showpacs]{revtex4}
\usepackage{amssymb}
\usepackage{graphicx}
\usepackage{amsmath}
\usepackage{times}
\usepackage{color}
\usepackage{subfigure}
\usepackage{setspace}
\usepackage{bm}% bold math

\begin{document}

\newcommand {\beq} {\begin{equation}}
\newcommand {\eeq} {\end{equation}}
\newcommand {\bqa} {\begin{eqnarray}}
\newcommand {\eqa} {\end{eqnarray}}
\newcommand {\da} {\ensuremath{d^\dagger}}
\newcommand {\ha} {\ensuremath{h^\dagger}}
\newcommand {\adag} {\ensuremath{a^\dagger}}
\newcommand {\no} {\nonumber}
\newcommand {\ep} {\ensuremath{\epsilon}}
\newcommand {\ca} {\ensuremath{c^\dagger}}
\newcommand {\ga} {\ensuremath{\gamma^\dagger}}
\newcommand {\gm} {\ensuremath{\gamma}}
\newcommand {\up} {\ensuremath{\uparrow}}
\newcommand {\dn} {\ensuremath{\downarrow}}
\newcommand {\ms} {\medskip}
\newcommand {\bs} {\bigskip}
\newcommand{\kk} {\ensuremath{{\bf k}}}
\newcommand{\kp} {\ensuremath{{\bf k'}}}
\newcommand {\qq} {\ensuremath{{\bf q}}}
\newcommand{\nbr} {\ensuremath{\langle ij \rangle}}
\newcommand{\ncap} {\ensuremath{\hat{n}}}

\begin{abstract}
  We theoretically study the relaxation of high energy single particle
  excitations into molecules in a system of attractive fermions in an
  optical lattice, both in the superfluid and the normal phase. In a
  system characterized by an interaction scale $U$ and a tunneling
  rate $t$, we show that the relaxation rate scales as $\sim Ct
  \exp(-\alpha U^2/t^2)$ in the large $U/t$ limit. We obtain explicit
  expressions for the exponent $\alpha$, both in the low temperature
  superfluid phase and the high temperature phase with pairing but no
  coherence between the molecules.  We find that the relaxation rate
  decreases both with temperature and deviation of the fermion density
  from half-filling. We show that quasiparticle and phase degrees of
  freedom are effectively decoupled within experimental timescales
  allowing for observation of ordered states even at high total energy
  of the system.

%discuss the implications of these results for
 % realization of many-body states of fermions in optical lattices.

% The Hubbard Hamiltonian  is believed to be a minimal model capable of explaining  high temperature  superconductivity, but so far  its phase diagram  resisted a definite solution.   Ultra-cold fermionic atoms trapped in optical lattices  provide a clean  implementation of the  Hubbard Hamiltonian and therefore  it is highly desirable to use these tunable  systems as quantum simulators to  yield insight on the Hubbard Hamiltonian  phase diagram.   In this work we describe a novel approach  for  controllable   preparation and  detection  of  d-wave superfluidity  in these systems using  an array of plaquettes created  via 2D optical superlattices. .
\end{abstract}

\title{Relaxation of Fermionic Excitations in a Strongly Attractive Fermi Gas in an Optical Lattice}

\author{  Rajdeep Sensarma$^{1}$, David Pekker$^{2}$, Ana Maria Rey$^3$, Mikhail D. Lukin$^4$  and 
Eugene Demler$^4$}  
 \affiliation{$1$ Condensed Matter Theory Center, University of Maryland, College Park, MD 20742,USA.\\
$2$ Department of Physics, California Institute of Technology, Pasadena, CA 91125, USA.\\
$3$ JILA, NIST and Department of Physics, University of Colorado, Boulder, CO 80309, USA.\\
$4$ Physics Department,
Harvard University, Cambridge, MA 02138, USA.
} 
\maketitle

Ultracold atoms on optical lattices~\cite{BlochReview} can be used for
simulating strongly interacting quantum many body
systems~\cite{Greiner,Esslinger,Bloch} with tunable Hamiltonian
parameters like interaction strength. Although the main focus of cold
atom experiments has been to obtain the equilibrium phase diagram of
various models, ultracold atomic systems also provide an unique
platform to study the intrinsic non-equilibrium dynamics of strongly
interacting many body systems. They are almost completely
decoupled from external environment and their low energy-scales lead
to long non-equilibrium time-scales over which it is possible to follow
the system without ultra-fast probes.

The issues of non-equilibrium relaxation dynamics are becoming an
important consideration for the state of the art cold atom
experiments~\cite{Weiss,Zoller,Hofferberth,DoublonRelax:Expt,Greiner2010,Bloch:expansion,Chin}
as well as recent pump-probe experiments with electron
systems~\cite{electron_expt}.  The tunability of Hamiltonian
parameters to access strongly interacting regimes is one of the
central attractive features of cold atoms. However, an implied
assumption in connecting the results obtained on optical
lattices to the physics of condensed matter systems is that the atoms
on the optical lattice have achieved thermal equilibrium at low
temperatures after tuning the parameters. Hence, it is important to
understand the relaxation dynamics and associated equilibration
timescales~\cite{Rigol,Altman,Kehrein,Lewenstein,Gritsev,DoublonRelax:Theory,Rosch2010,sc_noneq}
of these systems.

The attractive (or negative $U$) Hubbard model on optical
lattices is a lattice implementation~\cite{negU:Expt} of BCS-BEC
crossover~\cite{BlochReview,bcsbec}, which is a paradigm for
understanding strongly interacting superfluids. At weak coupling, this
model exhibits BCS superfluidity with an exponentially small critical
temperature. At strong coupling, the physics is governed by formation
of tightly bound molecules which undergo Bose-Einstein condensation at
low temperatures.  In this Letter, we consider the relaxation dynamics
of the attractive Hubbard model in a cubic lattice in the strong
coupling limit, where most fermions are paired to form molecules. We
focus on the decay of excess unpaired fermions present in the system
(either due to an external drive like lattice modulation or due to
sweeping of the Hamiltonian parameters) to form molecules. For these
high energy excitations, energy conservation requirements lead to a
very slow decay rate that scales super-exponentially with the ratio of
the interaction strength to the bandwidth of the system. Using a
particle-hole transform to map this problem to that of spin mediated
decay of double occupancies in the repulsive Hubbard model, we compute
the decay rate both in the low temperature superfluid phase and in the
high temperature paired phase for arbitrary filling fractions in the
lattice. We find that the decay rate decreases both with temperature
and with the deviation of the fermion density from half filling on
either side. We discuss the implications of these results for
maintaining adiabaticity during a sweep of Hamiltonian parameters.
 
We consider the one band attractive Hubbard model for fermions on a 3D cubic
optical lattice 
\beq 
H=-t\sum_{\nbr}\ca_{i\sigma}c_{j\sigma}-U\sum_i
n_{i\up}n_{i\dn} 
\eeq 
where $t$ is the tunneling matrix and $U$ is the local attraction
between the fermions. In the strong coupling limit (large $U/t$), the
fermions are paired to form tightly bound bosonic molecules with a
large binding energy $\sim U$, which undergo Bose condensation at a
temperature $\sim J=4t^2/U$, controlled by the kinetic energy scale of
the molecules.  The separation of the energy scales $U\gg t\gg J$
allows us to consider two different temperature regimes where most of
the fermions are paired into molecules: (a) the low temperature ($T
\ll J$) superfluid phase, where the molecules are Bose condensed and
(b) the high temperature ($T\sim t \gg J$) phase where the molecules
do not have phase coherence. We calculate the decay rate of unpaired
fermions in these two regimes.

The simplest process, where two unpaired fermions hop on top of each
other to form a molecule, is forbidden unless the binding energy of
the molecule ($\sim U$) is carried off by other excitations in the
system. There are two different modes of excitations where this excess
energy can be dumped: (a) kinetic energy of other unpaired fermions,
(with a scale $\sim t$) and (b) kinetic energy of the molecules, with
a scale $\sim J$. In this paper, we assume the later process is
dominant, which limits the density of unpaired fermions to be less
than $J/t \sim t/U$.
%   We first
% present a scaling argument for the super-exponential dependence of the
% molecular kinetic energy mediated relaxation of the unpaired
% fermions. 
Since the kinetic energy has an energy scale of $J=4t^2/U$,
$n \sim U/J \sim U^2/t^2$ molecular excitations have to be created to
release the binding energy. The matrix element for this process,
within $n^{th}$ order perturbation theory, is given by
\beq M \sim t\frac{t}{J}
\frac{t}{2J}.... \frac{t}{nJ}=\frac{t}{n!}\left(\frac{t}{J}\right)^n\sim
Ct \exp \left[-\alpha\frac{U^2}{t^2}\ln (U/t)\right] 
\eeq 
where we
have used $n!=n^n$ for large $n$ and $nJ=U$ to write the final form.
The decay rate $\Gamma\sim M^2$ thus decreases super-exponentially with $U/t$.
\begin{table}[t]
\caption{Equivalence of different quantities under the mapping between the attractive and the repulsive Hubbard model\label{table:mapping}}
%\begin{ruledtabular}
\medskip
\begin{tabular}{|c|c|}
\hline 
&  \\
Attractive Model & Repulsive Model \\
&  \\
\hline
Unpaired Fermions & Doublon hole pairs \\
\hline
Binding Energy & Mott gap\\
\hline
Deviation from half-filling & Magnetization \\
\hline
Superfluid order& Canted antiferromagnetic order \\
\hline
K. E. of Molecules & Superexchange energy \\
\hline
\end{tabular}
%\end{ruledtabular}
\end{table}

To obtain a physical picture of the decay process, it is instructive
to use a particle-hole transformation~\cite{Auerbach} which maps the attractive
(negative $U$) Hubbard model to a repulsive (positive $U$) Hubbard
model.  The attractive model with zero magnetization (equal up and
down spin densities) at any density is equivalent to the repulsive
model at half-filling (one particle per site) with a finite
magnetization proportional to the deviation of the fermion density in
the attractive model from half-filling, i.e. $m=(1/2)(1-\rho)$, where
$\rho$ is the fermion density in the attractive Hubbard model. Under
this transformation, molecule formation is mapped to the formation of
a Mott insulator, and the unpaired fermions are equivalent to the high
energy double occupancy (doublon) hole excitation, with the binding
energy of the molecules playing the role of the Mott gap. At
half-filling, the low energy physics of the repulsive Hubbard model
reduces to an antiferromagnetic Heisenberg model which exhibits a
canted antiferromagnetic order in its ground state in the presence of
finite magnetization. This spin ordering is equivalent to the
emergence of superfluidity in the attractive model with the kinetic
energy of the molecules playing the role of spin wave
fluctuations. The equivalence of the relevant quantities under the mapping is shown in Table.~\ref{table:mapping}.  
Here, we will use the language of
the spin model to look at quantitative estimates of the decay
timescales.

\begin{figure}[t!]
\includegraphics[width=0.4\textwidth]{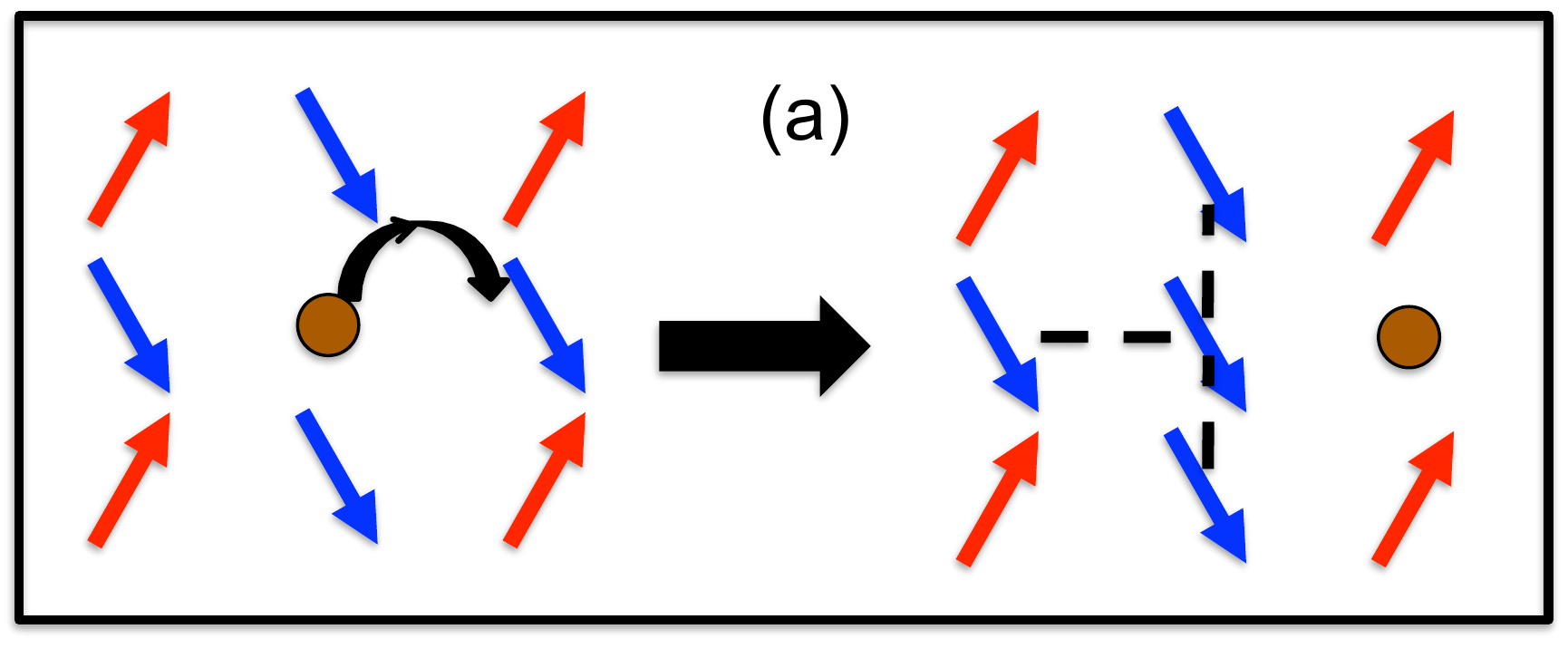}
\includegraphics[width=0.23\textwidth]{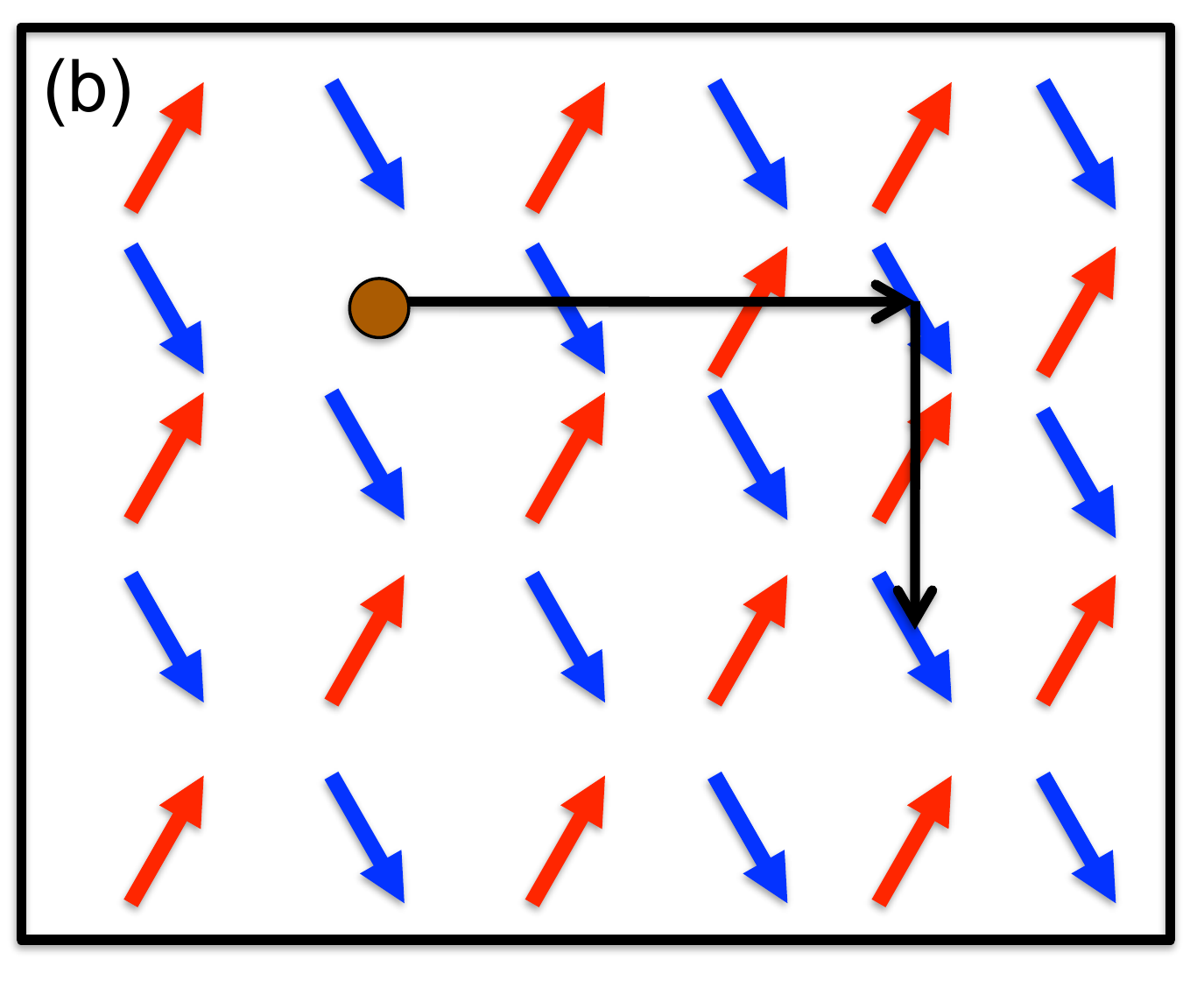}
\includegraphics[width=0.23\textwidth]{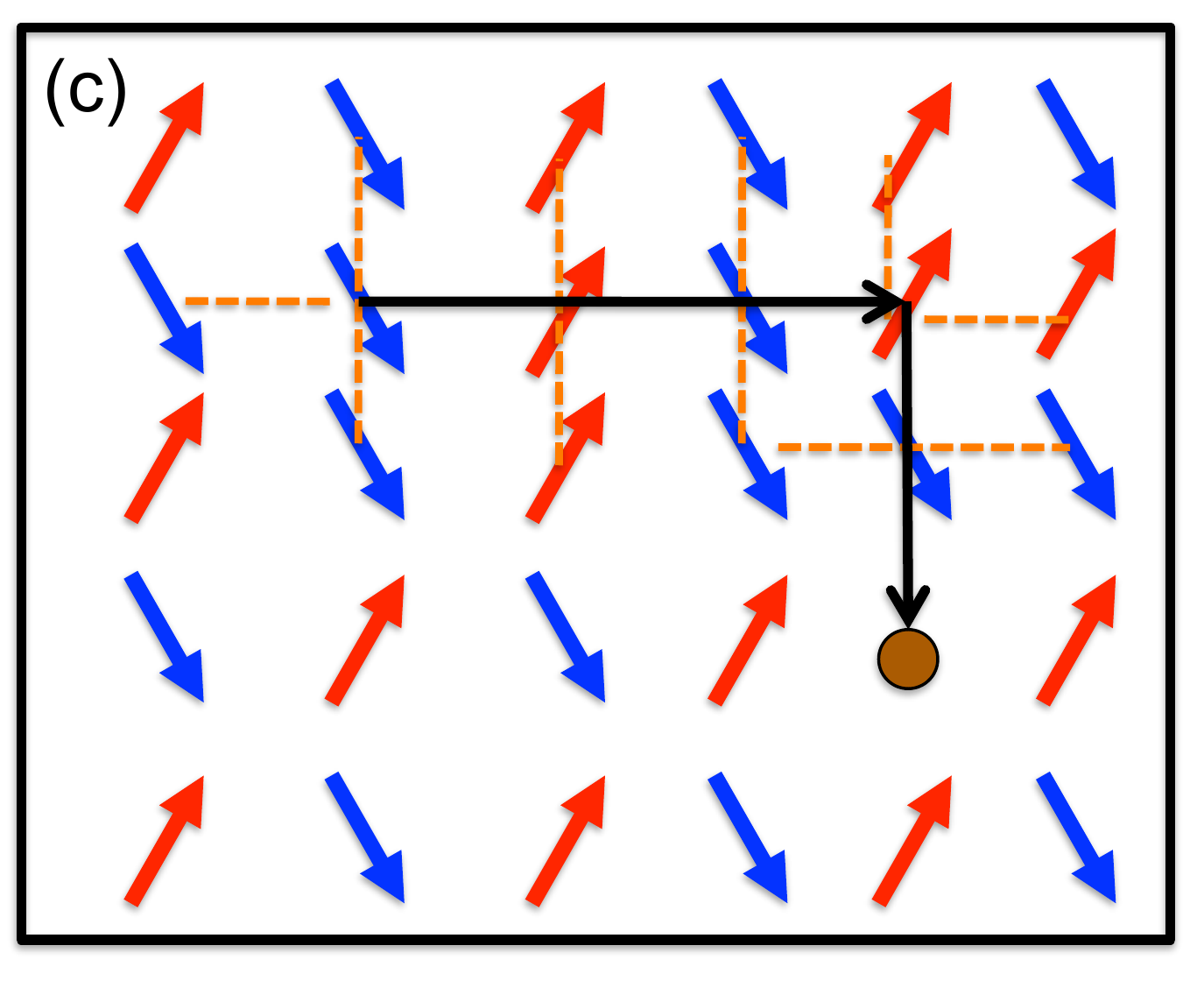}
\caption{ Hopping of a hole in a canted antiferromagnetic background.(a) A single hop moves back one spin, creating broken bonds shown by dashed lines. (b) and (c) Configurations before and after multiple hops of a hole. The solid line denotes the trajectory of the hole, while the dashed lines in (c) shows the broken bonds.}
\label{neel_hop}
\end{figure}

\begin{figure}[t!]
\includegraphics[width=0.4\textwidth]{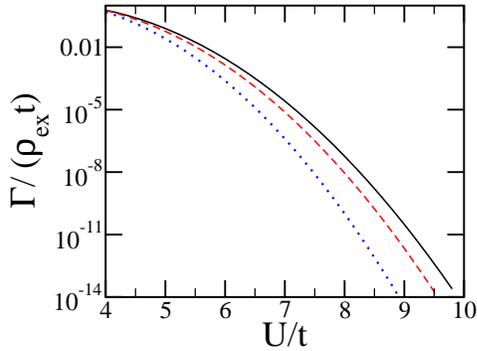}
\caption{The decay rate of unpaired fermions in the superfluid phase
  plotted as a function of $U/t$ for different fermion densities: (i) half-filling (solid black line) (ii) $\rho_f=0.7$ (red dashed line) and (iii) $\rho_f=0.5$ (blue dotted line).}
\label{fig:neeldecay}
\end{figure}

{\it Decay in the superfluid phase}: The superfluid phase of the
attractive fermions is represented by the canted antiferromagnetic phase
for the spins in the repulsive model. As a hole hops in the background
of a Mott insulator with canted spin ordering (shown
in Fig.~\ref{neel_hop}(a)) from a site $i$ to a neighboring site $j$,
it pushes back the spin on the site $j$ to the site $i$.  This
disrupts the spin texture and creates purely
ferromagnetic bonds between nearest neighbors, each of which gains an
energy of $(Jx/2)$, where $x=1-4m^2$ is proportional to the
antiferromagnetic component of the spin order.
 % In the language of the molecules, this is
% equivalent to the fact that the superfluid stiffness, which controls
% the molecular kinetic energy, decreases as one moves away from
% half-filling.
Hopping of the hole along a path creates
ferromagnetic bonds in the directions transverse to this path, thus
creating a domain wall in the system, as shown in
Fig.~\ref{neel_hop}(b) and (c). It is to be noted that hopping of
doublons in this background also leads to a similar process. In a cubic lattice each hop creates $z-2=4$ broken bonds. So, in order to accommodate
an energy $U$ and decay, the hole (doublon) need to traverse a path of
length $n=2U/[(z-2)Jx]$.
Within Fermi's golden rule, the decay rate is given by
\beq
\Gamma= 2\pi\rho_{ex}\nu(n) |M_{fi}|^2(n)
\eeq
where $\rho_{ex}$ is the density of holes (or equivalently of unpaired fermion excitations), and $M_{fi}(n)$ is the matrix element connecting the initial state
with a doublon and a hole to the final state with a domain wall of
length $n$ given by
 \beq 
M_{fi}(n)= \frac{t}{n!}\left[\frac{2t}{(z-2)Jx}\right]^n \sim t( t/U)^n
\eeq
Here $\nu(n)$, the density of
states, is proportional to the number of self-avoiding paths of length
$n$ connecting the doublon and the hole and is given by
\beq
\nu(n) =\int d^d r \frac{2n}{(z-2)Jx} S(n,r)G(r)
\eeq
where $S(n,r)$ is the number of self-avoiding paths of length $n$
connecting two points at a distance $r$ and $G(r)$ is the
dimensionless doublon-hole pair correlation
function. We assume that the doublons and holes are uncorrelated,
i.e. the pair distribution function is independent of $r$, or
$G(r)=1$. This assumption is valid in the limit of low density of
doublons, precisely the limit we are interested in. Then
$\nu(n) = \frac{2n}{(z-2)Jx}S(n)$, where $S(n)$, the total number of
self-avoiding paths of size $n$, scales as $\sim g^n n^k$ for large
$n$. The constants $g=4.68$~\cite{SelfAvoid:Sykes} and
$k=1/6$~\cite{SelfAvoid:Domb} for a cubic lattice has previously been
computed in the context of polymer physics. Using these, finally obtain the relaxation rate of unpaired fermions in the attractive Hubbard model
\beq
\Gamma \sim  \frac{t\rho_{ex}}{\sqrt{g}}\left(\frac{g}{8x}\right)^{2+k} \exp \left [-\left(\frac{U^2}{4xt^2}-3-2k\right) \ln \left(\frac{U}{\sqrt{g}t}\right)\right].
\eeq
Note that, for the attractive case, $x=2\rho-\rho^2$ is a measure of
the filling factor which vanishes both at $\rho=0$ (empty band) and
$\rho=2$ (completely filled band) and attains its maximum value at
half-filling ($\rho=1$). The low temperature decay rate, plotted as a
function of $U/t$ for different densities in Fig.~\ref{fig:neeldecay},
show the expected super-exponential scaling. As we move away from
half-filling (with increasing magnetization), the energy lost in a
single hop decreases, and longer domain walls are required to absorb
the excess energy, leading to a slower decay rate.

{\it Decay in the High Temperature Normal State}: We now consider the decay of the
single particle excitations in the high temperature phase ($T\sim
t\gg J$), where the fermions are still paired into molecules but there
is no superfluidity. In terms of the spin model, this regime
corresponds to a completely spin disordered phase with no spatial or
dynamic correlations. The single site Hilbert space can be occupied by an $\up$ spin with probability $1/2+m$ or by a down spin with probability $1/2-m$ (we are working at fixed magnetization). 
% %

As in the low temperature phase, the motion of holes (doublons) pushes
the spins along the trajectory by one site. However, the energy lost in
a given hop now depends on the configuration of the neighboring spins along the path. Since there
is no spatial correlation between the spins, the energy lost in each
hop can be treated as an independent random variable which takes the
values $Jr/2$ ($r=-4,-3,...4$), with the
probability $P(r)=(a/b)^{r/2}\sum^{4-r}_{i=0} \left._4C_i\right. \left._4C_{i+r}\right. a^ib^{4-i}$, where $\left._nC_k\right.$ is the Binomial co-efficient, $a=(1/2-m)^2$ and $b=(1/2+m)^2$. The
mean energy lost in each hop is $0$, while the variance of the
distribution is given by $J^2x/2$. The total energy lost in $l$ steps
is thus a Gaussian random variable with zero mean and a variance $lJ^2x/2$.
\beq
P(E,l) = \frac{1}{\sqrt{\pi xl}J}\exp[-E^2/(J^2lx)]
\eeq
Since the doublon needs to lose an energy $U$ to decay, one must now
average over decay processes from paths of length $l\geq n$ with the
probability distribution $P(U,l)$. The square of the matrix element
for a process involving $l$ hops is given by $M^2\sim
t^2(t^2/\ep_1^2)(t^2/\ep_2^2)...(t^2/\ep_l^2)\sim t^2 (2t^2/lJ^2x)^l$,
where $\ep_i$ is the energy lost after $i$ steps and, to leading
order, we have replaced $\ep_i^2$ by its average value $iJ^2x/2$. To
see the scaling in $t/U$, we note that the Gaussian probability
distribution $P(U,l)$ would be cut-off at a typical lengthscale
$l \sim 2U^2/(xJ^2)$, and hence the square of the matrix
element scales as $t^2(t/U)^{2l}$. Then, summing over all paths with
$l>n$ 
\beq 
\Gamma \sim 2\pi t^2 \rho_{ex}\int_n^\infty dl g^l
l^{k} P(U,l)\left(\frac{t}{U}\right)^{2l} 
\eeq
\begin{figure}[t!]
\includegraphics[width=0.4\textwidth]{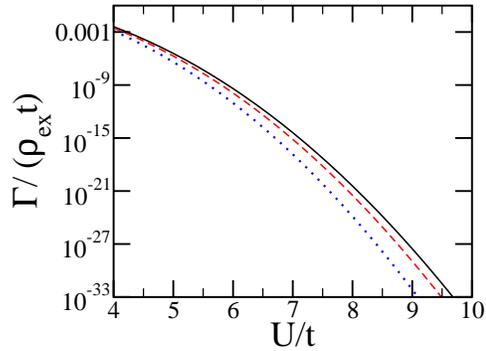}
\caption{The decay rate of unpaired fermions in the high temperature
  normal state plotted as a function of $U/t$ for different fermion densities: (i) half-filling (solid black line) (ii) $\rho_f=0.7$ (red dashed line) and (iii) $\rho_f=0.5$ (blue dotted line).}
\label{fig:htdecay}
\end{figure}
The high temperature decay rate, plotted as a function of $U/t$ in
Fig.~\ref{fig:htdecay}, is orders of magnitude smaller than the low
temperature decay rate. To understand this, note that the motion of
the high energy pair can both excite and de-excite low energy modes in
the background system. At zero temperature, these low energy modes are
unoccupied and the motion of the high energy pair then leads to
excitation of these modes. As temperature increases, the occupation
probability of the low energy modes increases and the motion of the
high energy pair randomly leads to excitation and de-excitation of
these modes. Thus the high energy pair loses its energy more
efficiently at lower temperatures and decays faster.

{\it Adiabaticity and sweep rates}: In cold atom experiments the
strongly interacting regime of model systems is accessed by tuning the
Hamiltonian parameters at a finite rate, which is limited by the
lifetime of the atoms in the trap. The tuning process
needs to be adiabatic in order to remain in the interesting low
temperature regime for the system. Since the microscopic relaxation
processes determine the timescale for equilibration, it is expected
that the relaxation timescales, along with experimental sweep rates,
would determine the limits of adiabaticity in these experiments. We
now make these ideas more precise by looking at the constraints due to
the slow decay of unpaired fermions (or equivalently doublon-hole
pairs).

We consider a system of attractive fermions at low enough temperatures
so that most of the fermions are paired to form molecules. In the
large $U/t$ limit, the density of unpaired fermions in equilibrium
$\rho_{ex}\sim \exp(-U/T)$, where $T$ is the temperature of the
system. We assume an adiabatic sweep of $U/t$ at a constant rate
$\gamma=\dot{(U/t)}$ and try to assess the limits where adiabaticity
fails. At these low temperatures, almost all the entropy comes from
the kinetic motion of the molecules; so for a constant entropy
process, we can assume $T/J=\lambda/4$  or $U/T= U^2/(\lambda t^2)$ along the sweep , where $\lambda$ is a constant. Now, adiabaticity will be maintained in the regime where
\beq
\dot{\rho_{ex}}=-\dot{(U/T)}\rho_{ex} =-\frac{2}{\lambda}\frac{U}{t}\gamma \rho_{ex}
\leq -\Gamma(U/t)\rho_{ex}
\eeq
 where $\gamma$ is the
experimental sweep rate. As the microscopic rate $\Gamma$ goes down
super-exponentially with $U/t$, this criterion would set an upper
limit of $(U/t)_{max}(\gamma)$, which is the maximum $U/t$ upto which
the system remains adiabatic when the parameters are swept with a rate
$\gamma$. 
% The same considerations apply for doublon density in a
% repulsive Hubbard model. In the quantitative analysis of the repulsive
% case one also needs to include the possibility of doublons moving to
% compressible edges and breaking apart there. While this relaxation
% mechanism involves an additional small probability of thermal
% activation to the edge, the relaxation rate in the compressible edge is
% only exponential in $U/t$, hence such process can be dominant. In this
% context, we note that the parabolic confinement potential present in
% real experiments favours the doublons at the center of the trap, while
% it favours migration of unpaired fermions to the edge of the
% trap. Hence the probability of transferring a doublon to the edge of
% the trap in the repulsive case will be different from the probability
% of transferring unpaired fermions to the edge in the attractive case
% and the precise equivalence would be broken. However, the
% spin-mediated decay does not involve transfer of particles to the edge
% of the trap and in this case the equivalence of molecule formation
% rate with the doublon decay rate is more precise
Our analysis shows that it is extremely difficult to keep the system
fully adiabatic in the strong coupling limit when either the tunneling
or the interaction are being changed, as the relaxation timescale of unpaired
atoms (or doublons for the repulsive case) can be anomalously
long. Experimentally this long timescale should manifest itself as a saturation in the molecular fraction with the saturation occurring at smaller values of $U/t$ for faster sweep rates.

At the same time, if relaxation of unpaired fermions is very slow
(longer then the timescale of experimental measurements), then they
can be considered as infinitely long lived and completely decoupled
from other degrees of freedom in the system like the phase
fluctuations of the superfluid order parameter. Similarly, in the
repulsive Hubbard model, if the goal is to observe antiferromagnetism,
one may worry that a small number of doublons can release enough
energy to destroy magnetic order. If the doublons are very long lived,
there will be a long timescale over which one can neglect relaxation
of doublons and analyze the quasi-equilibrium with "unbreakable"
doublons. Thus, within experimental timescales, there is an effective
spin-charge decoupling which makes it easier to observe spin ordering
even in the presence of high energy charge excitations. The idea of
realizing metastable states with long lived doublons has also been
discussed in the context of the $\eta$-paired state in the repulsive
Hubbard model~\cite{Rosch:eta}.

We have studied the decay of unpaired fermions in an attractive
Hubbard model. We have shown that the decay rate scales as $\sim Ct
\exp[-\alpha U^2/t^2]$ for large $U/t$ and computed the exponent
$\alpha$ both at low temperatures (superfluid phase) and high
temperatures (normal state of molecules). We find the decay rate
decreases with increase in both temperature and the deviation of the
fermion density from half-filling.
We also discussed implications of our analysis for realizing many-body
states in optical lattices. The downside of the long relaxation times
is that it is difficult to change parameters of the system fully
adiabatically. The upside of slow relaxation is that there is
effective decoupling of different degrees of freedom. So for example,
one may be able to achieve equilibration of phase (magnetic) degrees
of freedom, even when there is a finite density of unrelaxed single fermions (doublons).

The authors acknowledge fruitful discussions with A. Georges. RS
is supported by ARO-DARPA-OLE, ARO-MURI and JQI-NSF-PFC. AMR
acknowledges a grant from ARO with funding from DARPA-OLE. ED is
supported by CUA, AFOSR Quantum Simulation MURI, ARO MURI on
Atomtronics and NSF Grant No. DMR-0705472.

\end{document}